\documentclass[prl,twocolumn,superscriptaddress]{revtex4}
\usepackage{graphicx}
\usepackage{bm}

\begin{document}

\title{Charge mobility of discotic mesophases: a multiscale quantum/classical study}

\author{James Kirkpatrick}
\affiliation{Department of Physics, Imperial College London,
Prince Consort Road, London SW7 2BW, United Kingdom}

\author{Valentina Marcon}
\affiliation{Max-Planck-Institut f\"{u}r Polymerforschung,
  Ackermannweg 10, 55128 Mainz, Germany}

\author{Jenny Nelson}
\affiliation{Department of Physics, Imperial College London,
Prince Consort Road, London SW7 2BW, United Kingdom}

\author{Kurt Kremer}
\affiliation{Max-Planck-Institut f\"{u}r Polymerforschung,
  Ackermannweg 10, 55128 Mainz, Germany}

\author{Denis Andrienko}
\affiliation{Max-Planck-Institut f\"{u}r Polymerforschung,
  Ackermannweg 10, 55128 Mainz, Germany}

\date{\today}


\begin{abstract}
A correlation is established between the molecular structure and charge mobility of discotic mesophases of hexabenzocoronene derivatives by combining electronic structure calculations, Molecular Dynamics, and kinetic Monte Carlo simulations. It is demonstrated that this multiscale approach can provide an accurate {\em ab-initio} description of charge transport in organic materials.
\end{abstract}

\maketitle

Organic photovoltaic (OPV) and other organic optoelectronic devices rely upon blend films of two material phases, combining the attributes of a large interfacial area for charge separation and recombination and efficient vertical transport paths. Transport in organic electronic materials proceeds by incoherent hopping and depends upon local molecular ordering as well as on the presence  of percolation paths, and therefore is an intrinsically multiscale property. Improved mobilities are crucial to improved OPV device performance, and in turn require control of structure on both the intermolecular and macroscopic length scales.

Discotic liquid crystals could be ideal materials for OPV as they offer optimal
design possibilities for both the charge mobility and the
necessary underlying mesoscopic morphology. Discotic thermotropic liquid crystals
are formed by flat molecules
with a central aromatic core and aliphatic side
chains~\cite{ChandrasekharR90}. They can form columnar phases,
where the molecules stack on top of each other into columns, which
then arrange in a regular lattice. Along the stacks of aromatic
cores in the column one observes one dimensional charge
transport~\cite{vandeCraatsWFBHM99,Schmidt-MendeFMMFM01}.
Perpendicular to the column axis the charges have to tunnel
through the insulating side chains, resulting in a  much reduced
mobility. Therefore the discotics in a columnar phase act
as nano-wires and are a promising alternative to conjugated polymers,
where charge mobility in devices
is often limited by inefficient inter-chain hops.
Columnar mesophases
could, additionally, provide for laterally segregated phases of donor and
acceptor molecules with high interface areas and efficient
percolation pathways. The reported high charge
mobilities~\cite{vandeCraatsWFBHM99} would guarantee efficient
operation.
However, the
spatial arrangement of stacks is never perfect: the columns can be
misaligned, tilted, or form various types of topological defects,
which are metastable or even stable at ambient conditions.
In a recent molecular dynamics (MD) study we have shown that the degree
of order in discotic stacks of hexabenzocoronene (HBC) derivatives is sensitive
to the type of attached side chains~\cite{andrienko:2006.c}.
The local alignment of the molecules in columns and the global arrangement of the columns in
the mesophase, are thus both sensitive to molecular architecture.

A key question is the influence of local ordering in discotic mesophases on charge transport.
Computer simulation is a powerful tool for probing the influence of molecular structure on charge
transport.
However, the task is challenging because different theoretical methods are needed to describe
different length- and time-scale phenomena. Atomistic simulations are needed
for local molecular arrangements~\cite{andrienko:2006.c},
quantum chemical calculations are needed for the electron transfer mechanisms
(and interaction with the electrodes)~\cite{BredasBCC04}, and stochastic or rate-equation
dynamic methods for simulation of charge dynamics.
In this paper, we present the first statistical mechanics evaluation of the charge mobility based
on quantum chemical and atomistic MD methods. We
employ such a three level approach to explain the side chain dependence
of charge mobility for differently substituted HBC
derivatives.

Previous studies of liquid crystalline triphenylene derivatives
~\cite{KreouzisJCP01} have established that polaron hopping is an appropriate
description of microscopic charge transport in these materials. To use charge transport
theory as a predictive, rather than descriptive tool \emph{ab initio} calculations of
the rates for charge hopping between molecules are needed. To this end we use the
Marcus-Hush formalism~\cite{freed:6272} for the charge transfer rate $\omega_{ij}$
\begin{equation}
\omega_{ij}=\frac{|J_{ij}|^2}{\hbar} \sqrt{\frac{\pi}{\lambda kT}}
\exp \left[- \frac{(\Delta G_{ij} - \lambda)^2}{4\lambda kT} \right],
\end{equation}
where $J_{ij}$ is the  transfer integral for electron or hole transfer, $\Delta G_{ij}$ the
difference in free energy between the initial and final states, $\lambda$ the
reorganization energy, $\hbar$ Planck's constant, $k$ Boltzmann's
constant and $T$ the temperature. $J_{ij}$ and $\lambda$ can be
computed using quantum chemical methods~\cite{BredasBCC04}.   Note
that $J_{ij}$ is highly sensitive to the relative
position and orientation of the molecules involved, which will be determined,
in turn, using Molecular Dynamics.
If entropic effects and energetic disorder are ignored, $\Delta G_{ij} = {\bm F} \cdot {\bm d_{ij}}$,
where $\bm d_{ij}$ is the vector between the center of the molecules
involved in charge hopping and $\bm F$ the electric field.
In this study
we choose to neglect site energy differences due to electrostatic interactions between the molecules,
because such differences vanish when
the conjugated cores are parallel. Furthermore, the exclusion of site energy
disorder allows us to focus exclusively on the effect of configurational
disorder on charge transport, through disorder induced variations in $J_{ij}$.

Once all $\omega_{ij}$  are known, charge dynamics can be
simulated by Monte Carlo (MC) methods~\cite{BasslerJCP91} or by a
Master Equation (ME) approach~\cite{YuPRB01}. Though several
aspects of the problem are treated in the
literature~\cite{SenthilkumarJCP03,LemaurJACS04}, a
comprehensive study on large columnar systems does not yet exist. In
organic molecular crystals, Deng and coworkers~\cite{DengJPCB04}
showed that mobilities in pentacene can be reproduced by mixed
quantum chemical and molecular dynamics studies. Here we use a
similar combination of methods to achieve a truly \emph{ab-initio} description of
charge mobility in HBC: (i) molecular dynamics
simulations (MD) are performed on discotic mesophases of different
HBC derivatives, (ii) positions and orientations of molecules are
used to calculate the transfer integral $J$ for all neighboring
pairs of molecules in a column, and (iii) charge dynamics are
determined using both Master Equation and kinetic MC methods.

\begin{figure}
\begin{center}
\includegraphics[width=8cm]{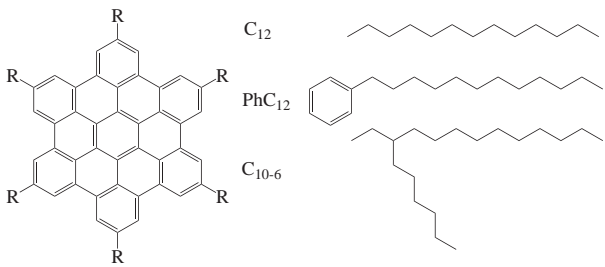}
\end{center}
\caption{Stick diagram of HBC derivatives with different side chains.
We studied alkyl chains of different lengths, ${\rm
C}_n$, with $n = 10, 12, 14, 16$; branched side chains, $\rm
C_{10-6}$; and the dodecylphenyl-substituted side chains $\rm PhC_{12}$.}
\label{fig:systems}
\end{figure}

All studied HBC derivatives are shown in
Fig.~\ref{fig:systems}. The molecules consist of a flat aromatic core and
six side chains. Experimental data on their structural
properties~\cite{FischbachPMFMSS02,vandeCraatsWFBHM99,FechtenkotterSHMS99,BrownSBMS99,HerwigKMS96},
as well as microwave conductivity (PR-TRMC)
data~\cite{vandeCraatsWFBHM99,pisula:2006.a} are available. In
some cases time-of-flight (ToF) mobility data are also
available~\cite{kastler:2006.a}.

\emph{Molecular dynamics.} In our simulation we adopt the united
atom approach for the side chains, and consider explicitly only
the hydrogen atoms belonging to the aromatic rings of the central
core. Details of the force field and the MD setup can be found in
Ref.~\onlinecite{andrienko:2006.c}. The molecules were arranged in
16 columns with 10, 20, 60, or 100 molecules in a column.
Production runs of $100\, \rm ns$ were performed at constant
pressure of $0.1\, \rm MPa$ and temperature $T = 300\,
\textrm{K}$ fixed using the Berendsen
method with anisotropic pressure coupling.
System configurations were saved every $10^4$ MD steps (the
timestep was $2\, \textrm{fs}$), yielding $200$ snapshots
for each run.

\emph{Quantum Chemical Calculations.} For each MD snapshot, the
overlap integrals $J_{ij}$ between intracolumnar nearest neighbors
$i$ and $j$ were calculated. To calculate $J_{ij}$, the aromatic core of
each molecule, as output by the MD simulation, is replaced by a rigid
copy of the energy minimized configuration, with the same axial and torsional
orientation as the MD result. Because of molecular symmetry,
the highest occupied molecular orbital (HOMO) and the lowest
unoccupied molecular orbital (LUMO) of HBC are doubly occupied.
Thus we have to calculate four transfer integrals for hole and
four for electron transport~\cite{SenthilkumarJCP03}.
For each component we used an adaptation of the ZINDO method, 
because of its speed and its insensitivity to polarization
effects~\cite{ValeevJACS06}. The effective $J$ values are
then taken as the root-mean-square of the four HOMO transfer
integrals for holes and of the four LUMO transfer integrals for
electrons~\cite{NewtonCR91}.
%
The inner-sphere reorganization energy for cation and anion radicals was
calculated using unrestricted wavefunctions, the
B3LYP functional and 6-31g** basis set. We found reorganization
energies $\lambda$ of $0.13\, {\rm eV}$ for cations and $0.11\,
\textrm{eV}$ for anions. The outer-sphere contribution was neglected.

\emph{Kinetic MC and Master Equation (ME) simulation.} For kinetic
MC, columns of molecules produced by MD are stacked periodically to produce a stack
$1\, \mu \textrm{m}$ thick. A uniform field is applied along the stack, a charge
carrier representing either an electron or a hole is introduced near one
end, and the charge drift is simulated by a continuous-time
random walk algorithm. The method used for MC simulation
was  similar to that in Ref.~\cite{Chatten05}
with adjustments for a disordered lattice.
The charge mobility $\mu$ is obtained from the transit time $t_{\rm tr}$ of the
simulated transient via $\mu=L/{F}t_{\rm tr}$ where $L$ is the stack thickness.
$t_{\rm tr}$ is taken as the point of intersection of the two
asymptotes to the simulated photocurrent plotted on a log-log plot, as would
be done in a ToF experiment.

To compare these to steady-state mobilities along the column, we
solved the linearized Master Equation~\cite{YuPRB01} for
continuous boundary conditions. The probability $P_i$
of a charge being on  site $i$ obeys the rate equation:
\begin{equation}
{\partial P_i } / {\partial t} = \sum \left[ \omega_{ij} P_j
 - \omega_{ji} P_i \right],
\end{equation}
where the sum is over neighbors $j \ne i $. $\mu$ is then obtained from the charge velocity
$
{\bm v} = \mu {\bm F} =\sum_{i \ne j} \left( \omega_{ji} P_i - \omega_{ij} P_j \right)
( {\bm r}_j - {\bm r}_i )
$, where ${\bm r}_i$ is the coordinate of site $i$ in the direction of ${\bm F}$. 

\begin{figure}
\begin{center}
\includegraphics[width=8cm]{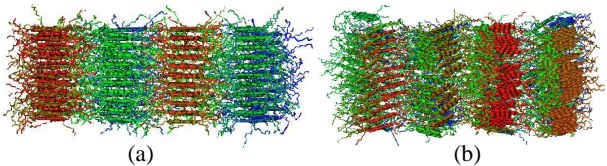}
\end{center}
\caption{
MD simulation snapshots of columns of 10 HBC molecules with (a) $\rm C_{12}$ and (b) $\rm C_{10-6}$ side chains at $300\,
\textrm{K}$. Columns are pre-arranged
on a rectangular lattice.}
\label{fig:snapC12}
\end{figure}

Representative snapshots of two of the systems are shown in
Fig.~\ref{fig:snapC12}. For $\rm C_{12}$, the aromatic cores are well aligned and
the columns are evenly spaced
whilst
for $\rm C_{10-6}$ the branched side chains lead to a more disordered columnar
structure, as studied in detail in Ref.~\onlinecite{andrienko:2006.c}. Here
we mention only the parameters relevant for charge transport,
namely, the intracolumnar molecular separation $h$ and nematic order parameter
$S$. The ordering for ${\rm C}_{10}$ -
${\rm C}_{16}$ side chains is almost perfect, while $\rm PhC_{12}$ and $\rm C_{10-6}$ are
slightly more disordered. For the bulkier $\rm C_{10-6}$ we find
 numerous defects in the columnar arrangement. The intracolumnar separations are practically the
same for all derivatives, as given in table~\ref{tab:systems}.

To study the effect of different side chains on mobility, 
we performed ToF simulations of photocurrent transients 
on columns constructed from copies of each MD snapshot and then averaged over
results for different snapshots. This sampling method is analogous
to the experimental ToF situation where charges are photogenerated in parallel
columns of different configuration, and the measured transient is the sum
of displacement currents from all columns. For the same ${\bm F}$, $t_{\rm tr}$
varies significantly from snapshot to snapshot, \emph{i.e.}
the mobility is sensitive to the particular
orientational and positional configurations of molecules in the column. Therefore,
detailed tests were performed to determine the number of different columnar
configurations as well as the length of the stacks
needed to obtain reproducible results. Tests showed that 100 different 
configurations are sufficient, and for 10 molecules per column the finite 
size error is less than $10 \%$.

\begin{figure}[htb]
\begin{center}
\includegraphics[width=8cm]{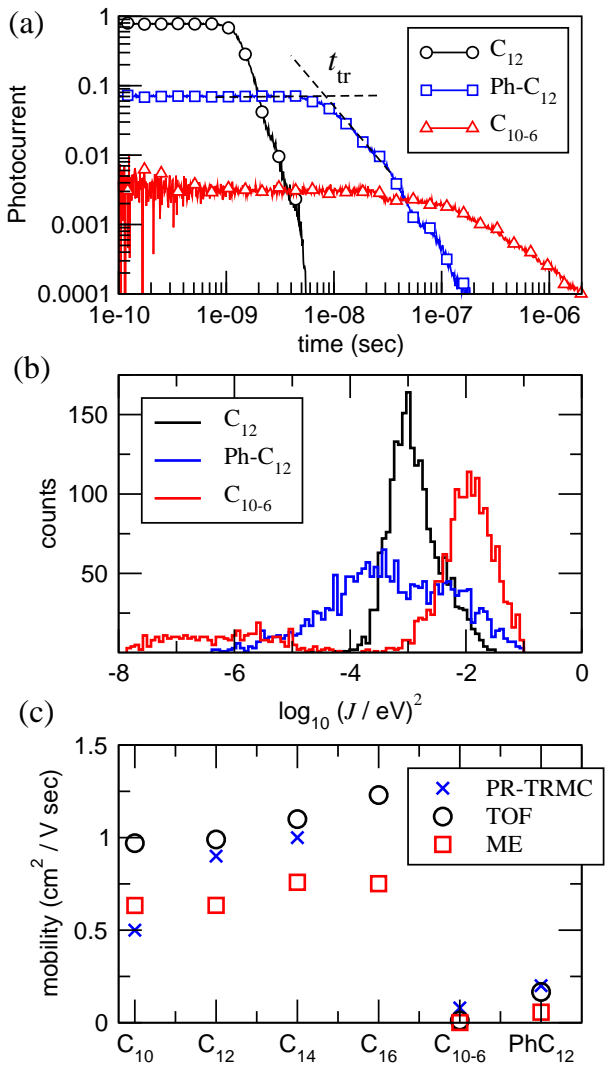}
\end{center}
\caption{ (a) Simulated ToF hole photocurrent transients for $\rm C_{12}$,
$\rm PhC_{12}$ and $\rm C_{10-6}$ 
at $T = 300\, {\rm K}$ and $F=10^5\, {\rm V cm}^{-1}$. 
Traces are  averaged over 200 snapshots with columns of length 10. 
(b) Frequency plots of the logarithm of the transfer integral squared. 
In each case, results for $\rm C_{10}$-$\rm C_{16}$ are practically indistinguishable from $\rm C_{12}$ on the scales used. (c) Sum of electron and hole mobilities calculated by ToF and by ME, in comparison
with experimental values. }
\label{fig:derivatives}
\end{figure}

The results of ToF simulations are shown in Fig.~\ref{fig:derivatives}a.
Comparison of the shape of the simulated transients for $\rm C_{10}$-$\rm C_{16}$ 
and $\rm C_{10-6}$ at the same $\bm F$ shows that the disordered $\rm C_{10-6}$ 
structures lead to relatively dispersive transients
while those for $\rm C_{10}$-$\rm C_{16}$ lead to the non-dispersive transients typical
of ordered materials~\cite {KreouzisPRB06}. In our approach, where energetic
disorder is neglected, disorder in charge transport can only arise from disorder
in the nearest-neighbour transfer integrals $J$, which in turn arises from disorder
in intermolecular separation and orientation. Figure~\ref{fig:derivatives}b compares distributions of $\log |J|^2$. As the data clearly show, the disordered compound leads to a much wider distribution of $J$ leading to a wider distribution of rates. In particular, the low $J$  peak represents bottlenecks
at discontinuities in the stacks, which act as traps in the one-dimensional
transport.

Finally, we compare our simulations to PR-TRMC measurements. 
PR-TRMC probes the sum of high frequency photoinduced conductivities due to both types of
charge carriers, and therefore probes the fastest contributions to charge transport, 
sensitive only to the local disorder within well ordered domains. 
PR-TRMC mobilities are therefore appropriate for comparison with our simulation since the column sizes
we are using are too small to include topological defects. ToF mobility data would,
in contrast, be heavily dependent on defects and on fluctuations with wavelengths
longer than the column size we have used.
\begin{table}
\begin{ruledtabular}
\begin{tabular}{llllllcc}
compound   & $\mu_{\rm PR-TRMC}$ & $\mu_{\rm ToF}^e$ & $\mu_{\rm ToF}^h$ & $\mu_{\rm ME}^e$ & $\mu_{\rm ME}^h $& $S$ & h \\ \hline
$\rm C_{10}$     & 0.5~\cite{vandeCraatsPCB98} & 0.22 & 0.75 & 0.14 & 0.49   & 0.98 & 0.36\\
$\rm C_{12}$     & 0.9~\cite{vandeCraatsPCB98} & 0.23 & 0.76 & 0.14 & 0.49   & 0.98 & 0.36\\
$\rm C_{14}$     & 1~\cite{vandeCraatsPCB98}   & 0.27 & 0.80 & 0.17 & 0.59   & 0.98 & 0.36\\
$\rm C_{16}$     & -                           & 0.29 & 0.91 & 0.17 & 0.58   & 0.98 & 0.36\\
$\rm C_{10-6}$   & 0.08~\cite{pisula:2006.a}   & -    & 0.01   & 6e-4 & 0.003 & 0.96 & 0.37\\
$\rm PhC_{12}$   & 0.2~\cite{vandeCraatsPCB98} & 0.036& 0.13 & 0.012 & 0.045 & 0.95 & 0.41
\end{tabular}
\end{ruledtabular}
\caption{Electron (e) and hole (h) mobilities ($\textrm{cm}^2\textrm{V}^{-1} \, \textrm{s}^{-1}$) of different compounds calculated
using time-of-flight (ToF) and master equation (ME) methods, in comparison
with experimentally measured PR-TRMC mobilities. Also
shown are the nematic order parameter $S$ and the average vertical
separation between cores of HBC molecules $h$ (nm).}
\label{tab:systems}
\end{table}
The simulated mobilities are compared with experimental data in table~\ref{tab:systems}
and in figure~\ref{fig:derivatives}c. The systematic overestimate of $\mu$ by the ToF compared to the ME simulation is due to the convention used here for $t_{\rm tr}$. Even though the differences
in experimental PR-TRMC mobilities are relatively small,
we are able to reproduce the experimental results not
only qualitatively but also quantitatively. This suggests that our
simulation procedure and the underlying morphologies are appropriate and
that the calculated transport parameters are all in the correct order of magnitude.
Of particular interest is the strong sensitivity of $\mu$ to
the degree of order in the columns, which is itself controlled by the side
chain type. The highest $\mu$ is obtained, in experiment and simulation,
for the linear side-chain derivatives which also present the highest degree
of intracolumnar order while the lowest $\mu$ is obtained for the branched side-chain derivative with
the least degree of order. 
The good correspondence between simulated and
PR-TRMC mobilities arises partly from the fact that both simulation and experiment
probe transport in relatively well aligned columns. To simulate ToF experiments,
large systems must be studied which will require a multiscale ansatz~\cite{hess:2006}.

In summary, we have  determined charge mobilities of several
derivatives of the discotic liquid crystal hexabenzocoronene. For the first time, three
methods were combined into one scheme: (i) quantum chemical
methods for the calculation of molecular electronic structures and
reorganization energies (ii) molecular dynamics for simulation of
the relative positions and orientations of molecules in a
columnar mesophase, and (iii) kinetic Monte Carlo simulations 
and Master Equation approach to simulate charge transport.
Applying this scheme to differently substituted HBC derivatives
we are able to reproduce the trends and magnitudes of mobilities
as measured by PR-TRMC. This study shows that it is possible to
understand and reproduce experimental charge transport parameters,
and, in future, accurately predict them.
This scheme is the outset for further studies, which will include
temperature dependencies as well as solid substrates to better
rationalize the self-organization of these systems. By employing
multiscale schemes realistic morphologies on a semi-macroscopic
scale should also become feasible.

\acknowledgments This work was partially supported by DFG via
International Research Training Group program between Germany and
Korea. V.M. acknowledges AvH foundation. J.K. acknowledges the EPSRC.
Discussions with K. M{\"u}llen and
J. Cornil are gratefully acknowledged.


\end{document}